\RequirePackage{lineno}
\documentclass{desyproc}
\usepackage{graphicx}
\usepackage{caption}
\usepackage{subcaption}

\begin{document}
\title{Search for Muonic Atoms at RHIC}

\author{{\slshape Kefeng Xin for the STAR Collaboration$^1$}\\[1ex]
$^1$Rice University, 6100 Main St MS-315, Houston, Texas, USA}

\contribID{39}

\confID{8648}  
\desyproc{DESY-PROC-2014-04}
\acronym{PANIC14} 
\doi  

\maketitle

\begin{abstract}
We present the search results for muonic atoms on $\sqrt{s_{NN}}=200$ GeV Au+Au collisions collected by the STAR experiment at RHIC. 
With the muon identification at low momentum, the invariant mass spectra were reconstructed. 
Clear signals are observed at the expected atom masses. 
Two particle correlations show that the production of the daughter particles happens at the same space-time point, presenting the signature of atom ionization. 
The fraction of primordial muons is extracted from $\pi$-$\mu$ correlations.
\end{abstract}

\section{Introduction}

Muonic atoms are like ordinary atoms except that the electrons are replaced with muons. These atoms have been studied in many fundamental physics experiments, such as precision measurements of proton size~\cite{protonSize} and nuclear quadrupole moments~\cite{quadrupole}. 
Muonic atoms with pions in the core have been produced from intense $K_l$ beam at Brookhaven National Lab \cite{pimuBNL} and Fermilab \cite{pimuFermi}. 
However hydrogen-like muonic atoms with more exotic particles in the core (kaons or antiproton) have never been observed. 
Heavy-ion experiments, with large amount of thermal muons and hadrons produced, make an ideal environment for the production of such exotic atoms. This provides us a great opportunity to make these discoveries. 

Muons that are involved in the atom production make the study particularly interesting in heavy-ion experiments, because thermal leptons are   considered to be ideal penetrating probes of hot QCD matter as their production rates rapidly increase with the temperature of the medium. 
However, one difficulty of measuring the thermal leptons is that they are mixed with a large amount of leptons from weak hadronic decays, which carry little information of the hot and dense matter. 
Muonic atoms are only produced by particles right after freeze-out, i.e.\ hadrons and thermal muons or muons from resonance decays like $\rho\rightarrow \mu^{+}\mu^{-}$, not by the muons from the weak hadronic decays, which are produced at a relatively late stage. 
Thus the idea of measuring the distributions of muonic atoms in heavy-ion collisions has been suggested by several theorists, Melvin Schwartz, Gordon Baym, Gerald Friedman, \cite{baym}, Joseph Kapusta, Agnes Mocsy, \cite{kapusta} etc.

\section{Analysis and Results}
The dataset used in this analysis is from Au+Au collisions at $\sqrt{s_{NN}}=200$ GeV collected by the STAR detector in year 2010. 
Central triggered events are selected to maximize the particle multiplicities. A total of 231 million events passed the event level selections. 
Particles are identified from the time-of-flight detector and the time projection chamber. 
The muon momentum is limited to 0.15-0.25 GeV/c to ensure the purity of the sample. 
The corresponding momenta for kaons and protons/antiprotons are 0.7-1.17 GeV/c and 1.33-2.22 GeV/c respectively. 

The invariant mass reconstruction is done with the combinatorial method. 
The combinatorial signal is constructed by pairing a hadron and a muon with opposite electric charges (unlike-sign method) from the same event. 
The background is constructed in two ways: a mixed-event method, in which a hadron and a muon with opposite electric charges from two different events are paired; and a like-sign method, in which a hadron and a muon with the same electric charge from a same event are paired. 

Note that the Coulomb effect becomes stronger when the two charged particles are close in phase space. 
In the unlike-sign method, two particles carry opposite charges, which produce attractive Coulomb force and thus enhance the mass distributions, especially at the low mass region. 
In contrast, in the like-sign method, the repulsive Coulomb force from the same charge suppresses the mass distributions at the low mass region. 
In the mixed-event method, there is no Coulomb effect for hadron-muon pairs. Therefore, the mixed-event backgrounds are used for acceptance correction of like-sign backgrounds: 
\begin{eqnarray}
LS_{+-}(corrected)=\sqrt{LS_{++}LS_{--}}\frac{ME_{+-}}{\sqrt{{ME_{++}ME_{--}}}},
\end{eqnarray}
where $LS$ and $ME$ stands for like-sign and mixed-event respectively, and the index stands for the charges for hadrons and leptons.  
Details of this correction are discussed in \cite{phenixDielectron} \cite{starDielectron}.

\begin{figure}[!hbt]
	\begin{subfigure}[b]{0.5\textwidth}
		\includegraphics[width=\textwidth]{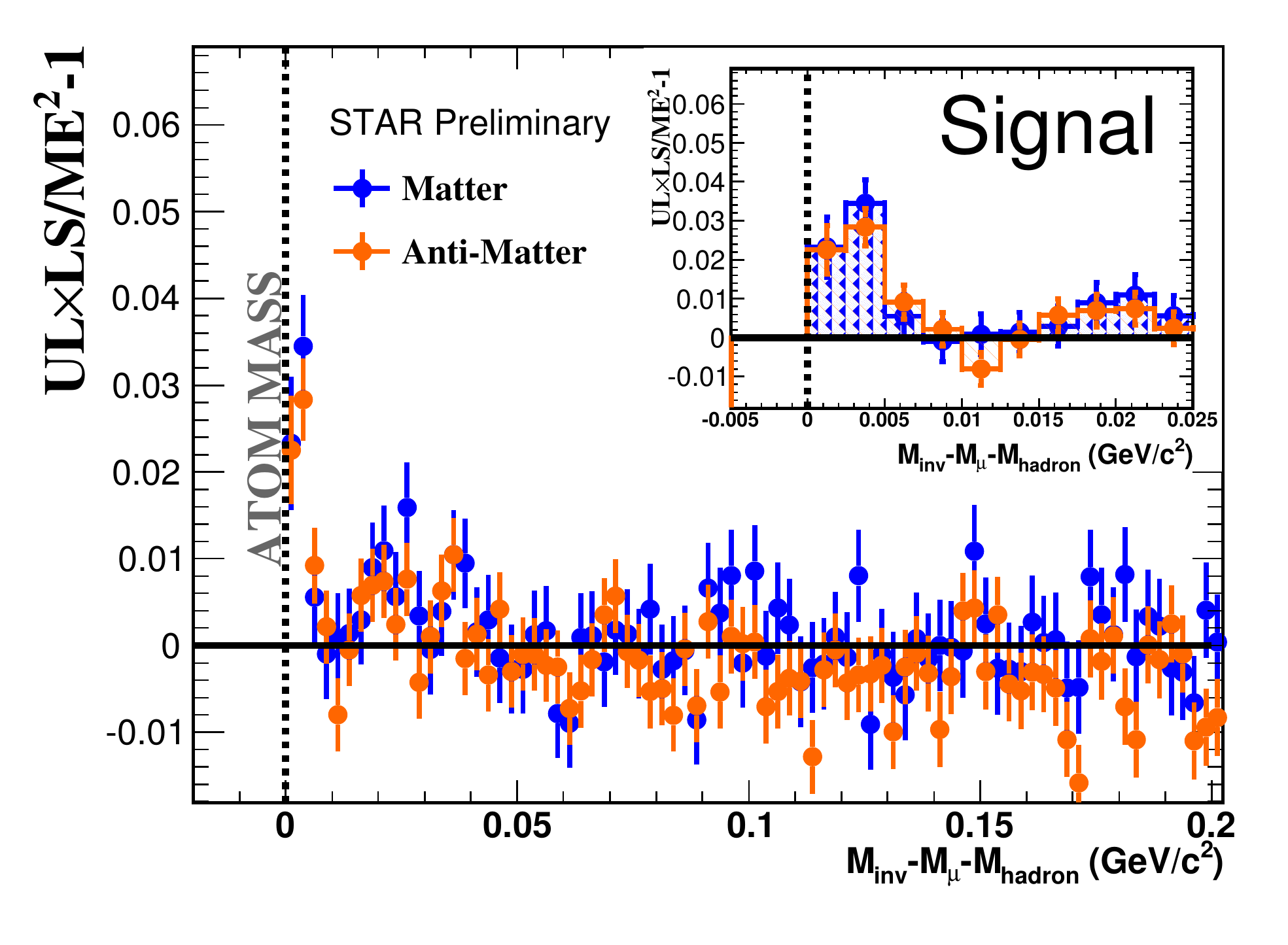}
		\caption{$p$-$\mu^-$ and $\bar{p}$-$\mu^+$ pairs.}
	\end{subfigure}
	\begin{subfigure}[b]{0.5\textwidth}
		\includegraphics[width=\textwidth]{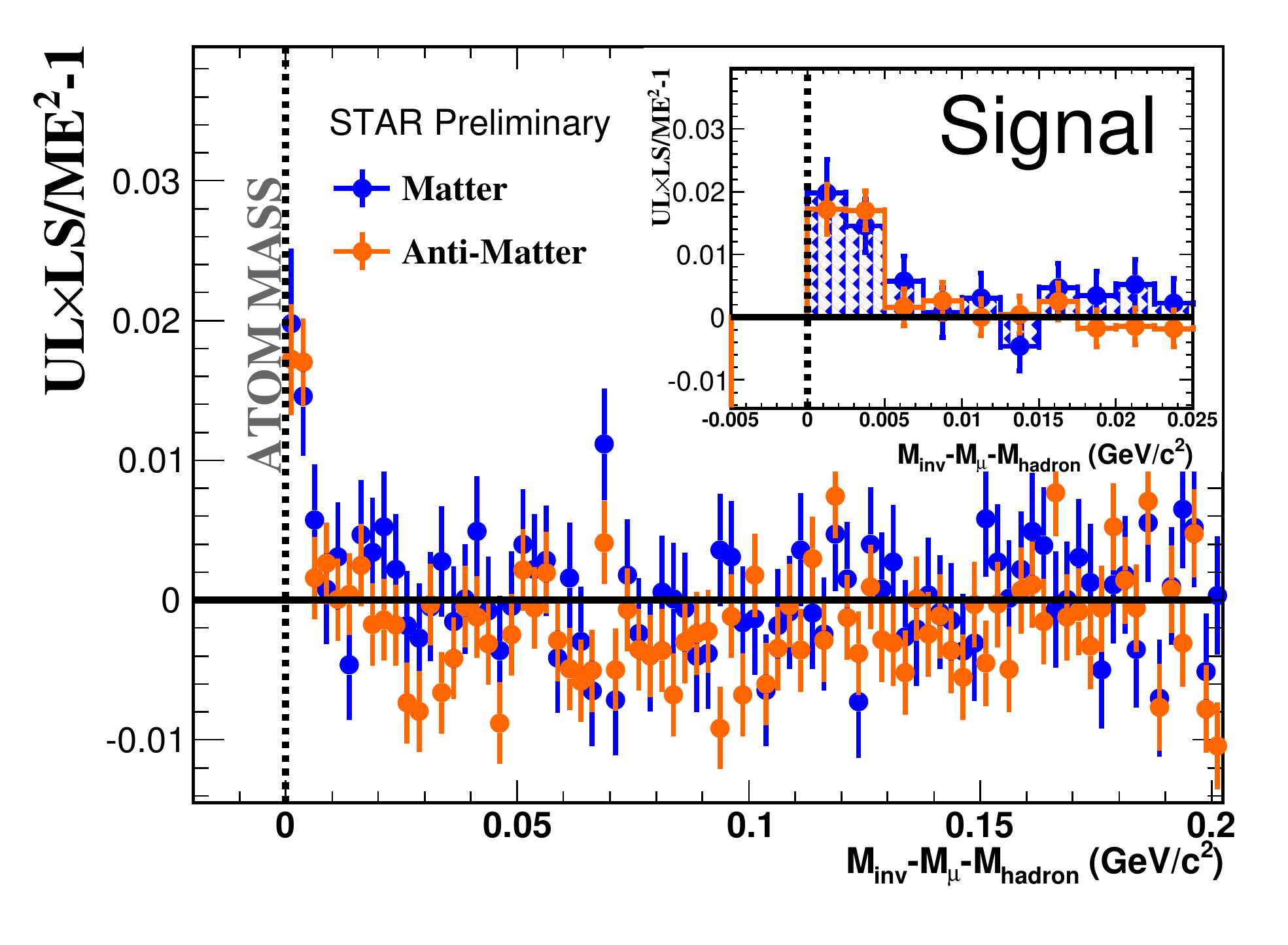}
		\caption{$K^-$-$\mu^+$ and $K^+$-$\mu^-$ pairs.}
	\end{subfigure}
	\caption{The pair invariant mass distributions of $UL\times LS/ME^2-1$ show peaks at the atom masses.}
	\label{Fig:mass}
\end{figure}

We adopted the following observable to cancel the trivial Coulomb effect and preserve the signal:
\begin{eqnarray}
UL\times LS/ME^2-1
\end{eqnarray}
where $UL\times LS$ stands for unlike-sign $\times$ like-sign, to cancel the Coulomb effect, and ME stands for mixed-event for normalization. 
After the rejection of the Coulomb force, we observe the sharp peaks at the expected zero net mass in Fig. \ref{Fig:mass}. 
The error bars show the statistical uncertainties. 
The signal is robust in both $K$-$\mu$ and $p$-$\mu$ systems and their antimatter systems. 

Femtoscopic correlations between two particles have also been used as a probe of muonic atoms.  
The correlation as a function of $k^*$, which is the magnitude of the momentum of either particle in the pair rest frame, shows how the interactions of the two particles change with respect to their distance in phase space. 
STAR has thoroughly studied the $K$-$\pi$ system ~\cite{femto}, in which only Coulomb interation dominates. 
For non-identified particles, a leading particle can be selected, and two cases can be distinguished by $C_{+}(k^*)$ and $C_{-}(k^*)$, which stand for the leading particle travels faster and slower, respectively.
Then the double ratio $C_{+}(k^*)/C_{-}(k^*)$ can be calculated to show the difference of the two cases. 
This method was successfully used in previous measurements to probe the space-time asymmetry of the emission of two particles \cite{femto}.
\label{sec:figures}
\begin{wrapfigure}{r}{0.55\textwidth}
  \begin{center}
    \includegraphics[width=0.55\textwidth]{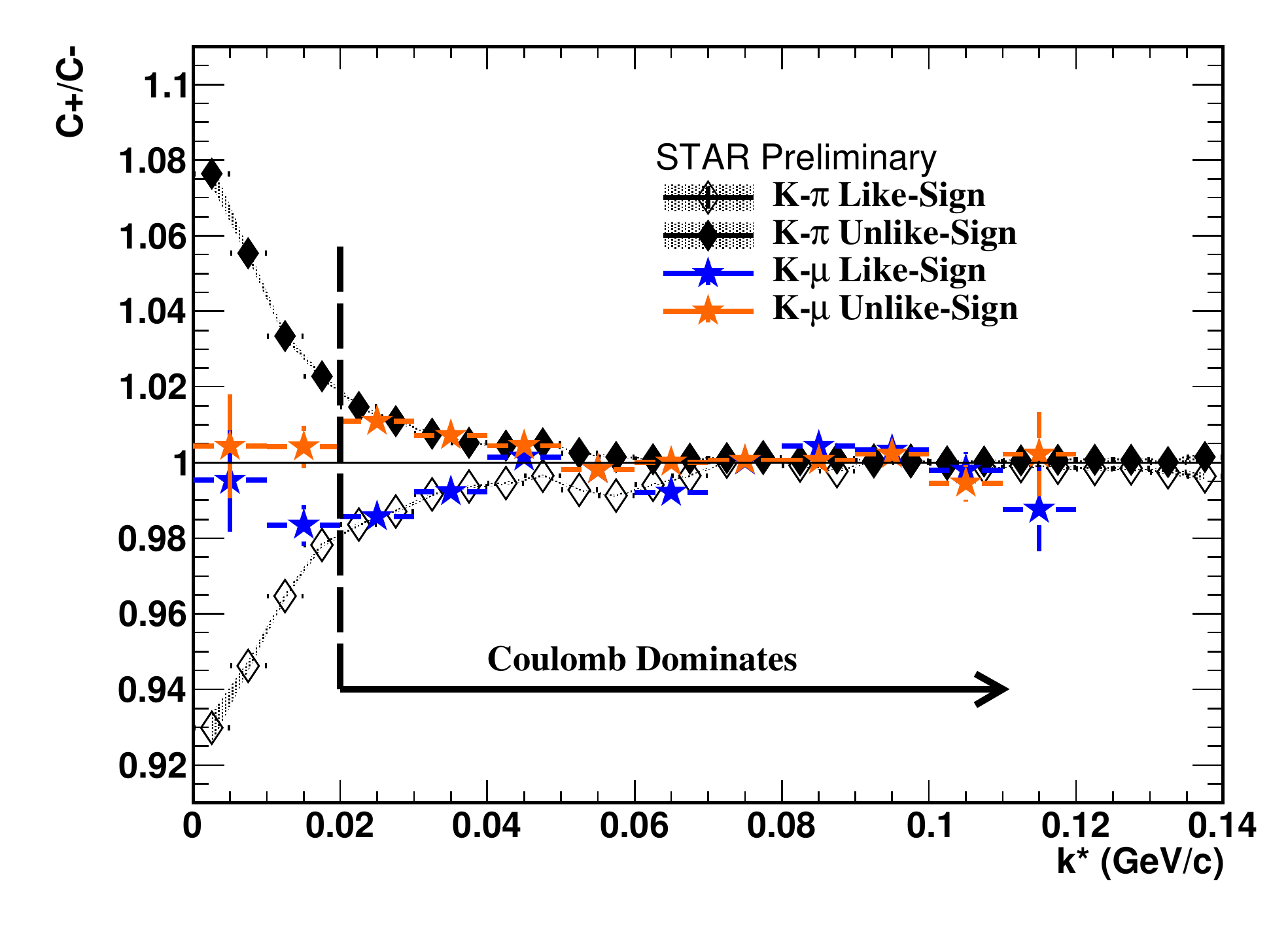}
  \end{center}
  \caption{The double ratio of the $K$-$\pi$ and $K$-$\mu$ systems show significant difference at low k*. The convergence to unity of $K$-$\mu$ suggests the ionization at the beam pipe after the production of muonic atoms.}\label{Fig:k4}
\end{wrapfigure}

The measurement of $K$-$\pi$ system is performed from the dataset and kinetic region similar as $K$-$\mu$ system which we used for muonic atom signal extraction. 
The origin of the non-unity in double ratio comes from the Coulomb interactions between the kaons and pions, which are later on enhanced in $C_+(k^*)$ and suppressed in $C_-(k^*)$ because of the space-time emission asymmetries of kaons and pions. 
The double ratio of $K-\mu$ system, overlaid on top of $K$-$\pi$ system, can be separated in two regions as shown in Fig.\ ~\ref{Fig:k4}. 
On the right of the dashed line, where only Coulomb interactions are expected in both systems, the double ratios of the two systems are consistent. 
This is consistent with the existence of the Coulomb force, which is a necessary condition to form muonic atoms. 
On the left of the dashed line, where the muonic atoms are expected to appear, when getting to very low $k^*$, instead of divergence, the double ratios of $K-\mu$ system show convergence to unity. 
The unity double ratio provides a signature of muonic atoms disassociation at the detector beam pipe, where the hadrons and the muons are separated from the bound state at the same space-time point. 

The $\pi$-$\mu$ correlations are also studied. 
A large amount of muons from weak decays can pass the track selections, and mix with the primordial muons. 
Thus these $\pi$-$\mu$ interactions inherit the interactions from $\pi$-$\pi$ interactions, which have two major sources, the electrostatic Coulomb interactions and quantum interference from identical pions~\cite{pionHBT}. 
The later factor generates a strong enhancement on the correlation functions. 
We denote the three correlation functions as the follows: 
\begin{itemize}
\item $A$ for correlations between pions and muons from simulated weak decays from real pions. 
\item $B$ for correlations between pions and inclusive muons, which is measured from data.
\item $C'$ for correlations between pions and primordial muons.
\end{itemize}

\label{sec:figures}
\begin{wrapfigure}{r}{0.55\textwidth}
  \begin{center}
    \includegraphics[width=0.55\textwidth]{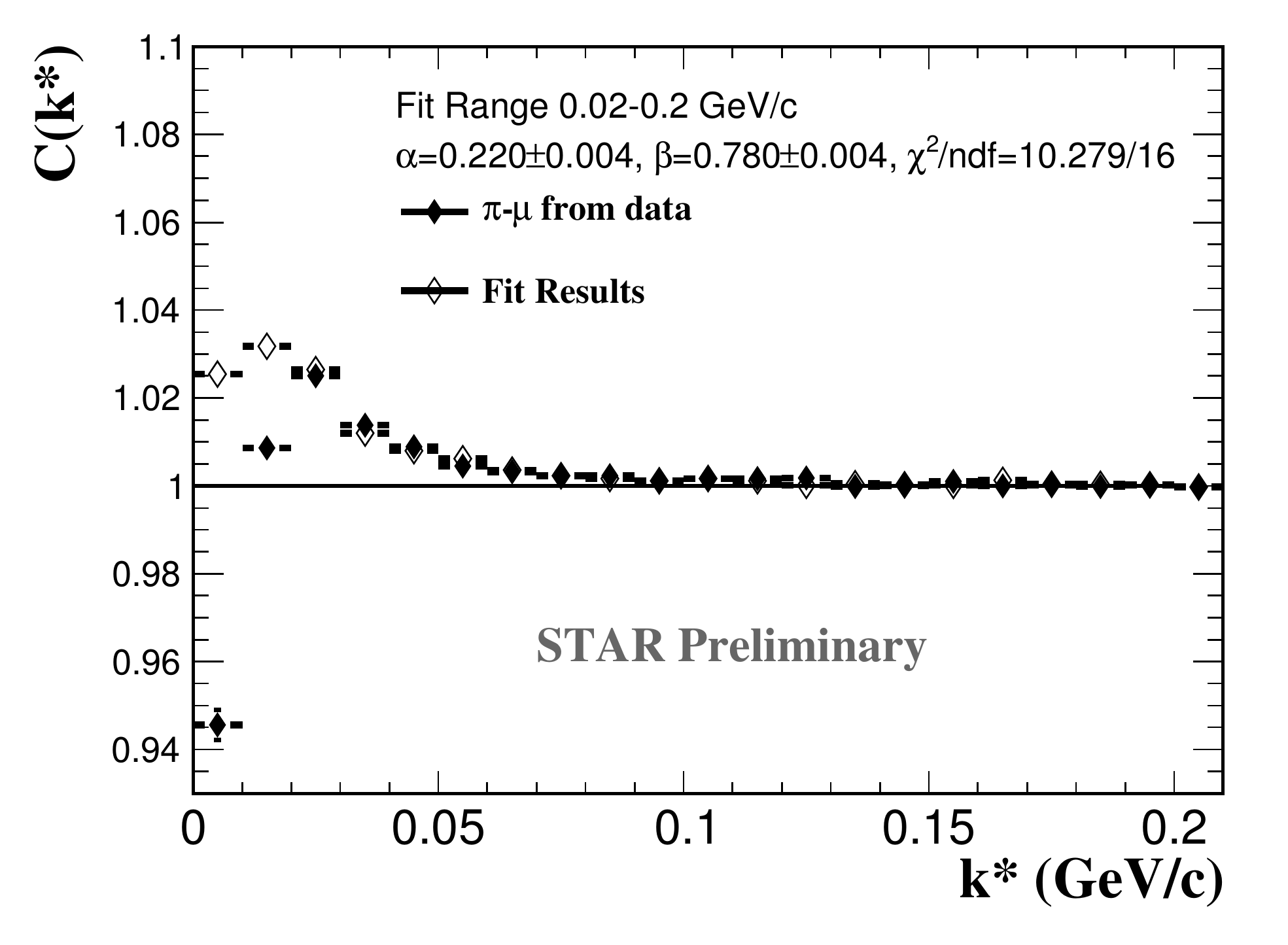}
  \end{center}
  \caption{Measured $\pi$-$\mu$ correlation function, fitted by $\pi$-$\pi$ correlation function and simulated $\pi$-$\mu_\mathrm{prim}$.}\label{Fig:fraction}
\end{wrapfigure}

The three functions satisfy the linear relationship: 
$B = \alpha\times C' + \beta \times A$, where $\alpha$ stands for the fraction of primordial muons from inclusive muons produced from the collisions.
$C'$ is then estimated by $\pi$-$\pi$ correlations, because of the fact that pion mass and muon mass are fairly close. 
To avoid quantum statistics enhancement, the correlation function from Coulomb between like-sign pairs is estimated from reversed unlike-sign pairs $C$. 
The relation then becomes:
$B = \alpha \times 1/C + \beta \times A$. 
The minimum $\chi^2$ fitting is performed in Fig.\  \ref{Fig:fraction}. 
If two particles have similar trajectories and orientation, implying that they are close in momenta space, the detector will not be able to have enough spacial resolution to distinguish them and will merge the two tracks. 
The fitting range is selected between 0.02-0.2 GeV/c, discarding the very low $k^*$ where the missing track problem is significant. 
The fitting results show that the fraction of primordial muons is 22.0$\pm$0.4\%.

\section{Conclusions}
Au+Au collisions at $\sqrt{s_{NN}}=200$ GeV collected by the STAR experiment are used in this measurement to search for muonic atoms.
The invariant mass distributions show clear signals at the expected mass position for $K^+$-$\mu^-$, $K^-$-$\mu^+$, $p$-$\mu^-$, and antiproton-$\mu^+$. 
The signal is robust after long-range Coulomb effect is rejected. 
The double ratio of $K$-$\mu$ system indicates the kaons and muons that are very close in phase space are emitted at the same space and time, which is consistent with muonic atom ionization. 
The fraction of primordial muons is extracted from the correlation method.

%
%
%
%
%
%


\begin{footnotesize}



%

\end{footnotesize}


\end{document}